# ABOUT THIOL DERIVATIZATION AND RESOLUTION OF BASIC PROTEINS IN TWO-DIMENSIONAL ELECTROPHORESIS

Sylvie Luche [1*], Hélène Diemer [2*], Chistophe Tastet [1], Mireille Chevallet[1], Alain Van Dorsselaer[2], Emmanuelle Leize-Wagner[2] and Thierry Rabilloud [1†]

1: CEA- Laboratoire de Bioénergétique Cellulaire et Pathologique, EA 2943, DRDC/BECP CEA-Grenoble, 17 rue des martyrs, F-38054 GRENOBLE CEDEX 9, France

2: Laboratoire de Spectrométrie de Masse Bio-Organique, UMR CNRS 7509, ECPM, 25 rue Becquerel, 67008 Strasbourg Cedex

(Running title): cysteine blocking for basic proteins

*: these authors contributed equally to this work

†: to whom correspondence should be addressed

Correspondence :
Thierry Rabilloud, DRDC/BECP
CEA-Grenoble, 17 rue des martyrs,
F-38054 GRENOBLE CEDEX 9
Tel (33)-76-88-32-12
Fax (33)-76-88-51-87
e-mail: Thierry.Rabilloud@cea.fr

Abbreviations:
DTDE: dithiodiethanol

Abstract

The influence of thiol blocking on the resolution of basic proteins by two-dimensional electrophoresis has been investigated. Cysteine blocking greatly increased resolution and decreased streaking, especially in the basic region of the gels. Two strategies of cysteine blocking were found efficient, either by classical alkylation with maleimide derivatives of by mixed disulfide exchange with an excess of a low molecular weight disulfide. The effect on resolution was important enough to allow correct resolution of basic proteins with in-gel rehydration on wide gradients (e.g. 3-10 and 4-12), but anodic cup loading was still required for basic gradients (e.g. 6-12 or 8-12). This shows that thiol-related problems are not solely responsible for streaking of the basic proteins on two-dimensional gels.

# 1 Introduction

One of the major weak points in proteomics using 2D gel electrophoresis lies in the difficulty in the analysis of basic proteins. Although considerable progress has been made recently [1-4] poor resolution of basic proteins, coupled with extensive horizontal streaking, is still frequently experienced. This has been attributed, at least in part, to cysteine oxidation at the basic part. In the classical IPG setup, the gel is rehydrated in a buffer containing a reducing agent such as DTT. As DTT is a weak acid (pKa close to 9), this means in turn that DTT migrates out of the basic part of the gel, leaving the proteins with the protection of the reducing agent and thus able to form mixed disulfides (intra or intermolecular). This oxidation is probably promoted by the electrochemical oxidation reactions taking place at the basic, cathodic electrode. In order to prevent these noxious phenomena, a strategy providing continuous influx of DTT from paper "reservoir" has been proposed [5]. However, this system is rather cumbersome and difficult to optimize from one sample to another, as it is based on the dynamic influx of reducer over the IPG run. A better strategy, in this case, would be to covalently block the cysteines, thereby preventing any noxious oxidation reaction such as disulfide formation or cysteine overoxidation. Cysteine blocking can be achieved either irreversibly by alkylation or reversibly by exchange with a large excess of a disulfide. The two strategies have been evaluated previously. Cysteine alkylation with either iodoacetamide or acrylamide has been proposed [6, 7]. Alternatively, blocking by disulfide exchange with dithiodiethanol [8] has also be proposed and shown to be useful. However, the real efficiency of both approaches has not been thoroughly investigated.

We therefore decided to investigate various methods of cysteine blocking in order to determine the role of cysteines in the resolution of the gels and, if important, to devise alternative and/or optimized methods for cysteine blocking.

# 2. Materials and methods

## 2.1. Sample preparation

In order to enable easy cysteine blocking, the sample must be reduced but contain no thiol. Thus, we devised an alternative buffer for whole cell extraction. The packed cells were first suspended in a minimal volume of 10mM Tris-HCl, pH 7.5, 250mM sucrose. The suspension was transferred into an ultracentrifuge tube. Then 4 suspension volumes of concentrated lysis buffer (8.75 M urea, 2.5 M thiourea, 5% CHAPS, 10mM Tris carboxyethyl phosphine and 20mM spermine base) were added. After 30 minutes at room temperature, the nucleic acids were spun down at 200000g for 30 minutes. The protein concentration was then estimated with a Bradford-type protein assay. 0.4% carrier ampholytes (final concentration) were then added and the sample stored at -80°C before use.

Alternatively, cellular subfractions with very low content in nucleic acids (e.g. mitochondria or microsomes) were used. They were diluted in 7M urea, 2M thiourea, 4% CHAPS, 0.4% ampholytes (3-10 range) and 4mM tributyl phosphine (prediluted at 100 mM in tetramethylurea). All the concentrations given above are final concentrations.

Cysteine blocking was performed by 2 alternative strategies. In the first one, the reduced sample is alkylated with 20 mM of alkylating agent at various pH. Carrier ampholytes of various pH ranges (2 pH units-wide) are used as buffers, and the pH is supposed to be the one of the midpoint of the pH range.To simplify the protocol, the maleimide derivatives are made as 10-fold concentrates in water (maleimide) or in dimethyl formamide (maleic hydrazide, N-methyl and N-ethyl maleimides). Acrylamide was used from a stock water solution, while the other alkylating agents (methylvinylsulfone, methyl methanesulfonate and methyl trifluoromethane sulfonate) were prepared as 200mM stock solutions in dimethyl formamide (i.e. 10 x concentrated). The reaction is left to proceed for 6 or 24 hours at room temperature. The alkylated extracts are then either loaded in a cup, or mixed with the gel rehydration solution (7M urea, 2M thiourea, 4% CHAPS, 0.4% ampholytes) [9].

In the second strategy, the reduced sample is either cup-loaded, or mixed directly with the gel rehydration solution (7M urea, 2M thiourea, 4% CHAPS, 0.4% ampholytes, and 100 mM disulfide). Various disulfides were tested, including dithiodiethanol, dithiodiglycerol and dithiodipyridine. Dithiodiglycerol was prepared in the laboratory by oxidation of 2 equivalents of thioglycerol with 1 molar equivalent of hydrogen peroxide.

## 2.2. Gel electrophoresis

Various immobilized pH gradients were tested, including a 3.75-10.5 linear pH gradient interpolated from a previously published 3-10.5 pH gradient [10], a non linear 4-12 pH gradient [1] and a linear 6-12 pH gradient [2]. The IPG gels were cast with plateaus at both sides [11]. The samples were applied either by in-gel rehydration or by cup loading, with the cup positioned at the acidic plateau. The IPG strips were then focused for 50000 Vh.

After focusing, the gels were equilibrated, with or without an iodoacetamide thiol alkylating step [12] and then run on a second dimension 10%T gel.the gels were then silver stained with an ammoniacal silver protocol [13] or, for subsequent mass spectrometry analysis, with a fluorescent ruthenium complex [14] or with colloidal Coomassie Blue [15].

## 2.3. Mass spectrometry

*In gel digestion :*

Excised gel slice rinsing was performed by the Massprep (Micromass, Manchester, UK) as described previously [14]. Gel pieces were completely dried with a Speed Vac before digestion. The dried gel volume was evaluated and three volumes trypsin (Promega, Madison, US) 12.5ng/$\mu$l freshly diluted in 25mM NH4HCO3, were added. The digestion was performed at 35°C overnight. Then, the gel pieces were centrifuged for 5 min in a Speed Vac and 5$\mu$l of 35% H2O/ 60% acetonitrile/ 5% HCOOH were added to extracted peptides. The mixture was sonicated for 5 min and centrifuged for 5 min. The supernatant was recovered and the procedure was repeated once.

*MALDI-TOF-MS analysis*

Mass measurements were carried out on an ULTRAFLEX$^{TM}$ MALDI TOF/TOF mass spectrometer (Bruker-Daltonik GmbH, Bremen, Germany).This instrument was used at a maximum accelerating potential of 20kV and was operated either in reflector positive mode. Sample preparation was performed with the dried droplet method using a mixture of 0.5mL of sample with 0.5mL of matrix solution. The matrix solution was prepared from a saturated solution of α-cyano-4-hydroxycinnamic acid in H2O/ 50% acetonitrile diluted 3 times. Internal calibration was performed with tryptic peptides resulting from autodigestion of trypsin (monoisotopic masses at m/z=842.51 ; m/z=1045.564 ; m/z= 2211.105).

*MS Data analysis*

Monoisotopic peptide masses were assigned and used for databases searches using the search engine MASCOT (Matrix Science, London, UK). All human proteins present in Swiss-Prot were used without any pI and Mr restrictions. The peptide mass error was limited to 70 ppm, one possible missed cleavage was accepted.

## 3. Results

### 3.1. Evaluation of published procedures

We first investigated the influence of cysteine blocking by disulfide exchange [8], or by alkylation with acrylamide. In order to be fully efficient, these options require first the sample to be reduced, and then to react with a large excess of a low molecular weight disulfide or of acrylamide. As the alkylation process with acrylamide is pH and time dependent, we used two different times (6 and 24 hours) and three different pH (6.5, 8 and 9). Typical results are shown on figure 1. It can be easily seen that treatment with dithiodiethanol sharply increased resolution, while treatment with acrylamide induced a decrease resolution. In order to get more details on the molecular mechanisms at play, and especially to know whether the action of dithiodiethanol and acrylamide is on the proteins themselves or rather on the surrounding medium, we analyzed the separated proteins by peptide mass fingerprinting. Typical results are shown on figure 2 and Table 1. They clearly show that whatever the alkylation conditions, cysteine alkylation with acrylamide is far from being complete, while spurious alkylation on lysine begins to take place, giving rise to artefactual peptides. On the dithiodiethanol side, the proportion of underivatized cysteine at the end of the equilibration period is fairly low (less than 10%) and correlates with the good resolution observed.

### 3.2. Investigation of other alkylation processes

In order to obviate the spurious alkylation phenomena observed with acrylamide, two tracks were explored. the first one is to use an alkylating agent much more powerful than acrylamide, but at a lower pH to increase the selectivity for thiols over amines. To this purpose, we investigated alkylation with methylvinylsulfone, methyl methanesulfonate and methyl trifluoromethane sulfonate, at pHs ranging from 5 to 6 and in the presence of a large excess of amino groups (from carrier ampholytes used as buffers) to decrease the risk of amine alkylation. Typical 2D gels obtained after such an alkylation are shown on figure 3. The resolution is terrible, which suggests major problems linked to spurious alkylation on proteins amino groups. We thus followed the opposite track, which is to use a less reactive alkylating chemical, and we focused on the maleimide derivatives. We used the classical N-ethyl maleimide, but also maleimide itself, N-methyl maleimide or maleic hydrazide. Typical gels are shown on figure 4, and clearly demonstrate a good resolution. Here again, peptide mass fingerprinting was carried out to characterize the separated proteins. Typical results are shown on figure 5 and table 2. While it is quite clear that cysteine alkylation with maleimide derivatives is a rather inefficient process (compare with table 1), at least under our conditions, it appears sufficient to improve the resolution of the basic proteins. Furthermore, spurious alkylation on lysine is fairly uncommon (one case detected only), which may explain the increased resolution over acrylamide-alkylated samples.

### 3.3. Investigation of the disulfide exchange process

The disulfide exchange process has been introduced recently in a short but appealing paper [8]. From the data shown in this paper, the real mechanism of action of dithiodiethanol is rather unclear. We first sought to establish whether other disulfides would be equally efficient.
The only theoretical constraint on the disulfide is that it is electrically neutral to prevent spurious transport phenomena at the high disulfide concentration required by this approach. We thus tested three different disulfides, dithiodiethanol, dithiodiglycerol and dithiodipyridine. The first two are highly water soluble, the latter being soluble only up to 150mM. Typical results are shown on figure 6. A clear increase of resolution can be seen with all disulfides. As could be anticipated, dithiodipyridine produced a lower resolution in the acidic region, where the pyridine moiety begins to be protonated. Dithiodiethanol and dithiodiglycerol were equally effective and yielded a high resolution all over the pH range. This showed that the positive effect of disulfide does not depend on their structure, within the constraints that they must be neutral and water-soluble.

### 3.4. Pushing the limits of the system

In proteomics studies, the analysis of the minor proteins often requires that milligram protein quantities of the starting complex extract are loaded onto the IPG strip. While the alkylation process can be easily scaled up to face this challenge, this is not obvious for the disulfide exchange process, which requires a large excess of the low molecular weight disulfide over the protein thiols. This large excess might be not guaranteed with 100 mM disulfide at high protein loads, especially if by some unknown mechanism, some of this disulfide is consumed at the basic, reducing electrode. We therefore decided to test higher disulfide concentration, up to 0.5 M, i.e. 6% (v/v) in the gel. At such concentrations, any reduction alkylation process after IEF become unpracticable, and we thus performed a very simple equilibration in urea-glycerol-SDS buffer. Typical results are shown on figure 7, and corresponding mass spectra are shown on figure 8. It is easily seen that a correct resolution is achieved at 500mM dithiodiethanol. Furthermore, the cysteine-containing peptides are correctly seen on the MS spectra, but as a mercaptoethanol adduct, i.e. with a mass shift of 76 Da. Quantitative analysis of the spectra showed that the derivatization yield was above 85%. No improvement nor deterioration was seen when 1M dithiodiethanol was used (data not shown).
Last but not least, we then investigated whether the cysteine blocking by this method also increased the resolution in basic pH gradients up to the point that in gel rehydration could be used instead of cup loading. The results, shown on figure 9, show that it is unfortunately not the case, and that cup loading is still required for optimal resolution in the basic gradients.

### 4. Discussion
The importance of protein thiols in the resolution problems encountered in the basic range have been described for quite a while [5], but no real convenient solution had been described until recently. The DTT process [5] is difficult to control, as is the alkylation process with iodoacetamide or acrylamide [6]. The first user-friendly solution came from a recent paper introducing the thiol-disulfide exchange as a cysteine derivatization process. In order to get a better understanding of the molecular processes at play, we performed a combined study by 2D electrophoresis in gradients extending in the basic pH range, coupled with peptide mass fingerprinting to characterize the reactions taking place on the proteins.

Our results clearly demonstrate the danger of the alkylation procedures with acrylamide, which have often been proposed as a convenient alkylation process [6, 7]. Even at moderate pH (6.5) cysteine alkylation is not complete after 6 hours, as shown by the presence of underivatized cysteine, while alkylation already takes place on lysines. The same problem has been demonstrated previously for iodoacetamide [6], and we demonstrate here that it takes place for all reactive alkylating agents. The only alkylating agents which do not give rise to spurious alkylation are those derived from maleimide. However, their alkylation yield is poor under the conditions prevailing in isoelectric focusing solubilisation buffers (i.e. multimolar concentrations of urea and thiourea, as well as the presence of carrier ampholytes to scavenge spurious reactions on lysine and the presence of detergents). Although the determination of the cysteine derivatization by integration of the signals on the mass spectra is always questionable, we believed that the solubility and ionization properties were dictated mainly by the peptide backbone and only slightly modulated by the side group on the cysteine chain. This entitled us to perform comparative integration of the mass signals arising from the same peptide but varying in the side group grafted to the cysteine thiol. The only exception to this rule might be the underivatized peptide itself, which may crosslink and give rise to insoluble products. This means in turn that this signal may be underevaluated, thereby leading to an overevaluation of the real yield of the derivatization process.

Whatever the real yields are, it is puzzling to see that a substantiable increase in resolution can be reached with only 30% alkylation of the protein thiols, as shown by the maleimide experiments, and that increased blocking of the thiols does not really change the picture. Maybe complete and specific thiol blocking would even improve the situation. However, we believe that other phenomena than simple thiol-linked problems are at play, as demonstrated by previous studies [16].

On a less mechanistic but more practical point of view, we suggest to use routinely 0.5M dithiodiethanol in the gel for pH gradients extending above pH 8. This procedure has the additional benefit to simplify the equilibration procedure (one bath of urea-SDS-glycerol buffer) without compromising the subsequent resolution of the cysteine-containing peptides in mass spectrometry. Simpler treatment of the protein-containing gel plugs prior to digestion is even possible, as no reduction-alkylation is required. However, if special cysteine derivatization must be performed at this stage, the reversibility of the thiol-disulfide exchange still make it possible.

Legends to figures

**Figure.1**. cysteine blocking with acrylamide or dithiodiethanol
Bovine mitochondrial proteins were separated by 2D gel electrophoresis. First dimension: Linear pH 3.75-10.5 immobilized pH gradient . Equilibration after IPG by the DTT-iodoacetamide two-step method. Second dimension: 10%T gel. Detection with silver staining.
A: proteins are reduced in 50mM DTT and run in an IPG gel containing 20mM DTT. B: proteins are reduced in 5mM TBP and run in a gel containing 100mM DTDE. C: proteins are reduced with 5mM TBP and alkylated prior to IPG with 20 mM acylamide for 6 hours at pH 6. D: same as C, but alkylation is performed at pH 9.

**Figure. 2**: MALDI MS spectra of malate dehydrogenase
A: alkylated with 100 mM DTDE, equilibrated with the DTT-iodoacetamide method.
The peptides marked with a star correspond to unalkylated cysteine-containing peptides with m/z of 1281.7 (positions 92-104, Cys 93) , 1313.7 (positions 204-215, Cys 212), 1432.7 (positions 79-91, Cys 89), 1898.9 (positions 280-296, Cys 285). The peptides marked with a + correspond to the same peptides with a +57 Da shift, i.e. carboxamidomethylated with iodoacetamide. The corresponding m/z are 1338.7, 1370.7, 1489.7, 1955.9. The peptides marked with a † correspond to the cysteine-containing peptides with a 76 Da shift, i.e. with a mixed disulfide with a merccaptoethanol moiety. The corresponding m/z are 1357.8, 1389.8, 1508.8, 1975.0.

B: alkylated with 100 mM acrylamide at pH 9 for 6 hours, equilibrated with the DTT-iodoacetamide method.
The peptides marked with a star correspond to unalkylated cysteine-containing peptides with m/z of 1281.7, 1313.7, 1432.7, 1898.9. The peptides marked with a + correspond to the same peptides with a +57 Da shift, i.e. carboxamidomethylated with iodoacetamide. The corresponding m/z are 1338.7, 1370.7, 1489.7, 1955.9. The peptides marked with a † correspond to the cysteine-containing peptides with a 71 Da shift, i.e. alkylated with acrylamide. The corresponding m/z are 1352.7 1384.8, 1503.7, 1969.9. The peptides marked with an ◊ correpond spurious alkylation peptides, i.e. peptides alkylated on a lysine and therefore undigested by trypsin at this position. The corresponding m/z and positions are 1099.6 (298-307), 1127.6 (157-165), 1584.9 (315-328), 1724.9 (240-257), 2214.2 (158-176)

**Figure. 3**: cysteine blocking with highly reactive alkylating agents
Bovine mitochondrial proteins were separated by 2D gel electrophoresis. First dimension: Linear pH 3.75-10.5 immobilized pH gradient . Equilibration after IPG by the DTT-iodoacetamide two-step method. Second dimension: 10%T gel. Detection with silver staining.
A: proteins are reduced in 5mM TBP and run in a gel containing 100mM DTDE. B: proteins are reduced with 5mM TBP and alkylated prior to IPG with 20 mM methyl methane sulfonate for 3 hours at pH 6. C: proteins are reduced with 5mM TBP and alkylated prior to IPG with 20 mM methyl triflate for 3 hours at pH 6.  C: proteins are reduced with 5mM TBP and alkylated prior to IPG with 20 mM methyl vinyl sulfone for 3 hours at pH 6.

**Figure 4**: cysteine blocking with maleimide derivatives
Bovine mitochondrial proteins were separated by 2D gel electrophoresis. First dimension: Linear pH 3.75-10.5 immobilized pH gradient . Equilibration after IPG by the DTT-iodoacetamide two-step method. Second dimension: 10%T gel. Detection with silver staining.
A: proteins are reduced in 5mM TBP and run in a gel containing 100mM DTDE. C: proteins are reduced with 5 mM TBP and alkylated prior to IPG with 20 mM methylmaleimide for 6 hours at pH 6. D: same as B, but alkylation is performed with ethyl maleimide; D: same as B, but alkylation is performed with maleic hydrazide

**Figure. 5**: MALDI MS spectra of malate dehydrogenase
A: alkylated with methyl maleimide, equilibrated with the DTT-iodoacetamide method.
The peptides marked with a star correspond to unalkylated cysteine-containing peptides with m/z of 1281.7 (positions 92-104, Cys 93) , 1313.7 (positions 204-215, Cys 212), 1432.7 (positions 79-91, Cys 89), 1898.9 (positions 280-296, Cys 285). The peptides marked with a + correspond to the same peptides with a +57 Da shift, i.e. carboxamidomethylated with iodoacetamide. The corresponding m/z are 1338.7, 1370.7, 1489.7, 1955.9. The peptides marked with a † correspond to the cysteine-containing peptides with a +111 Da shift, i.e. alkylated with methyl maleimide. The corresponding m/z are 1392.7, 1424.7, 1543.7, 2009.9

A: alkylated with ethyl maleimide, equilibrated with the DTT-iodoacetamide method.
The peptides marked with a star correspond to unalkylated cysteine-containing peptides with m/z of 1281.7 (positions 92-104, Cys 93) , 1313.7 (positions 204-215, Cys 212), 1432.7 (positions 79-91, Cys 89), 1898.9 (positions 280-296, Cys 285). The peptides marked with a + correspond to the same peptides with a +57 Da shift, i.e. carboxamidomethylated with iodoacetamide. The corresponding m/z are 1338.7, 1370.7, 1489.7, 1955.9. The peptides marked with a † correspond to the cysteine-containing peptides with a +125 Da shift, i.e. alkylated with ethyl maleimide. The corresponding m/z are 1406.8, 1438.8, 1557.8, 2024.0

**Figure 6**: Cysteine blocking with organic disulfides
Bovine mitochondrial proteins were separated by 2D gel electrophoresis. First dimension: Linear pH 3.75-10.5 immobilized pH gradient . Equilibration after IPG by the DTT-iodoacetamide two-step method. Second dimension: 10%T gel. Detection with silver staining. The proteins are reduced by 5mM TBP prior to application on the IPG strip
A: IPG strip containing 100 mM DTDE; B: IPG strip containing 100mM dithiodiglycerol; C: IPG strip containing 100 mM dithiodipyridine

**Figure 7**: Cysteine blocking with high concentrations of dithiodiethanol
Bovine mitochondrial proteins (1mg) were separated by 2D gel electrophoresis. First dimension: Linear pH 3.75-10.5 immobilized pH gradient. Second dimension: 10%T gel. Detection with colloidal Commassie Blue. The proteins are reduced by 5mM TBP prior to application on the IPG strip
A: IPG strip containing 100 mM DTDE. Equilibration after IPG by the DTT-iodoacetamide two-step method; B: IPG strip containing 500mM DTDE. Equilibration without DTT nor iodoacetamide

**Figure 8** MALDI MS spectra of proteins treated with 500mM DTDE. Single step equilibration without DTT nor iodoacetamide

A: malate dehydrogenase
The peptides marked with a star correspond to unalkylated cysteine-containing peptides with m/z of 1281.7 (positions 92-104, Cys 93) , 1313.7 (positions 204-215, Cys 212), 1432.7 (positions 79-91, Cys 89), 1898.9 (positions 280-296, Cys 285). The peptides marked with a † correspond to the cysteine-containing peptides with a 76 Da shift, i.e. with a mixed disulfide with a merccaptoethanol moiety. The corresponding m/z are 1357.8, 1389.8, 1508.8, 1975.0.

B: ATP synthase, alpha subunit
The peptides marked with a star correspond to unalkylated cysteine-containing peptides with m/z of 1270.7 (positions 242-252, Cys 244) and 3331.6 (positions 271-301, Cys 294) (Not detected). The peptides marked with a † correspond to the cysteine-containing peptides with a 76 Da shift, i.e. with a mixed disulfide with a merccaptoethanol moiety. The corresponding m/z are 1346.8 and 3407.7

**Figure 9**: Influence of application method for analysis of basic proteins
Bovine mitochondrial proteins (0.1 mg) were separated by 2D gel electrophoresis. First dimension: Linear pH 6-12 immobilized pH gradient. The proteins are reduced by 5mM TBP prior to application on the IPG strip, which contains 0.5 M DTDE. Equilibration without DTT nor iodoacetamide Second dimension: 10%T gel. Detection with silver staining.
A: application by in gel rehydration; B: application by cup loading at the anodic side

Table 1: Cysteine alkylation yield with acrylamide and dithiodiethanol

| alkylation with acrylamide | | | alkylation with dithiodiethanol | | |
|---|---|---|---|---|---|
| +0 Da | +57 Da | +71 Da | +0 Da | +57 Da | +76 Da |
| 0.12 | 0.48 | 0.40 | 0.08 | 0.78 * | 0.14 |

The alkylation yield is calculated by integrating the mass spectrometry peak areas for the unalkylated masses, the +57 Da adduct (alkylated with iodoacetamide), the +71 Da adduct (alkylated with acrylamide) and the +76 Da adduct (alkylated with dithiodiethanol). The statistics are carried out on all the cysteine-containing peptides for 8 proteins (ATP synthase alpha subunit, Isocitrate dehydrogenase (NADP dependent), creatine kinase, malate dehydrogenase, aspartate aminotransferase, succinate dehydrogenase, NADH ubiquinone oxidoreductase (B22 and PDSW subunits).

*: the high yield of iodoacetamide adduct is due to the reversibility of the alkylation with dithiodiethanol under the equilibration conditions chosen

Table 2: Cysteine alkylation yield with maleimide derivatives

| alkylation with N-methyl maleimide | | | alkylation with N-ethyl maleimide | | |
|---|---|---|---|---|---|
| +0 Da | +57 Da | +111 Da | +0 Da | +57 Da | +125 Da |
| 0.17 | 0.53 | 0.30 | 0.17 | 0.45 | 0.38 |

The alkylation yield is calculated by integrating the mass spectrometry peak areas for the unalkylated masses, the +57 Da adduct (alkylated with iodoacetamide), the +111 Da adduct (alkylated with N-methyl maleimide) and the +125 Da adduct (alkylated with N-ethyl malerimide). The statistics are carried out on all the cysteine-containing peptides for 8 proteins (ATP synthase alpha subunit, Isocitrate dehydrogenase (NADP dependent), creatine kinase, malate dehydrogenase, aspartate aminotransferase, succinate dehydrogenase, NADH ubiquinone oxidoreductase (B22 and PDSW subunits).

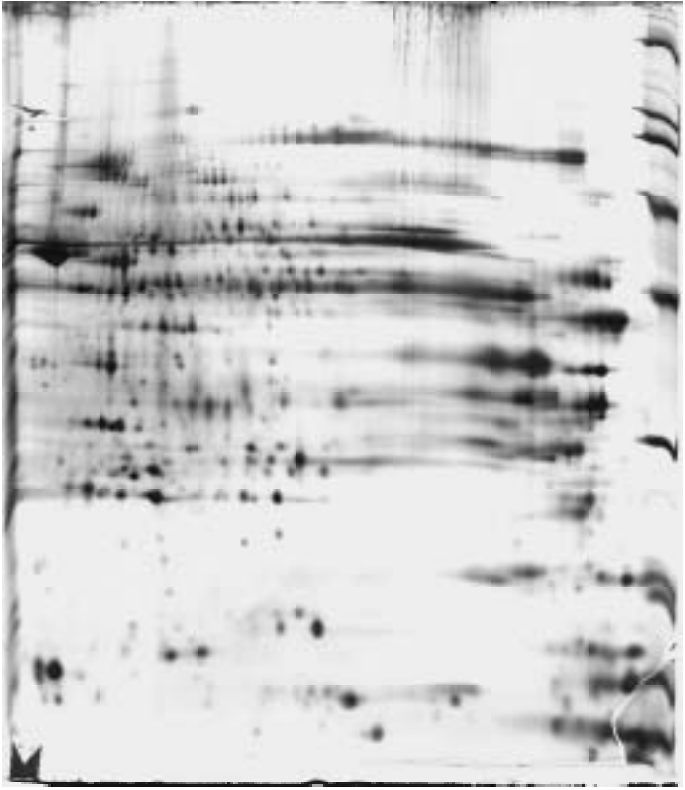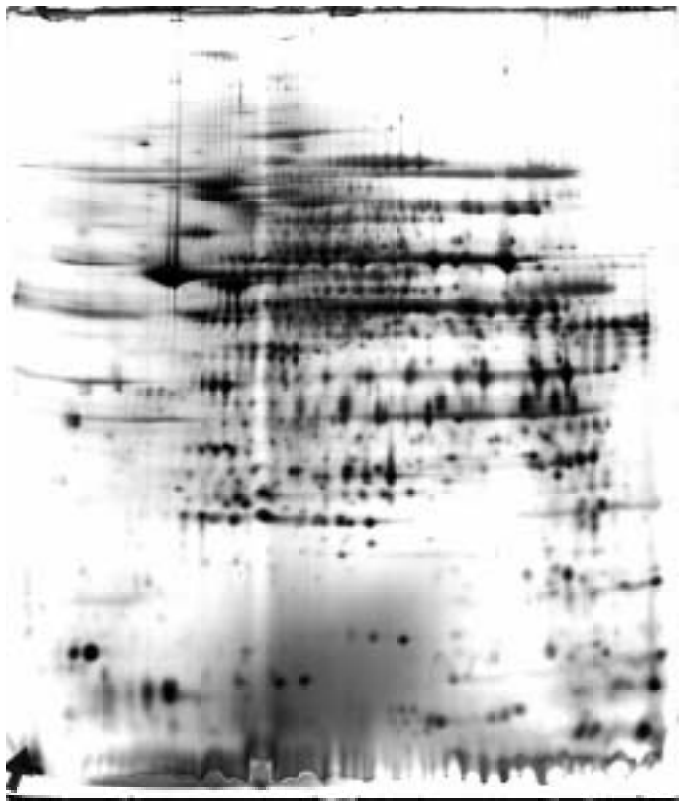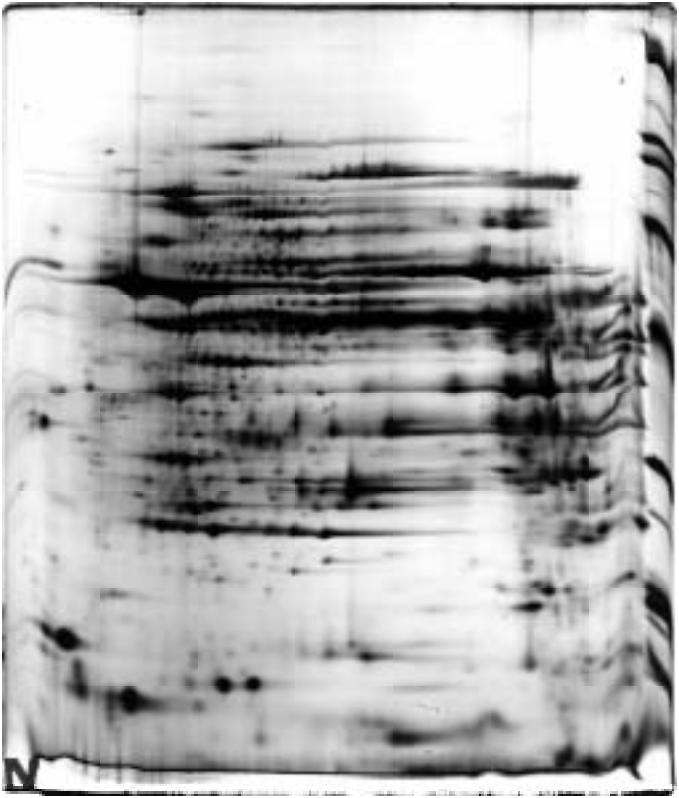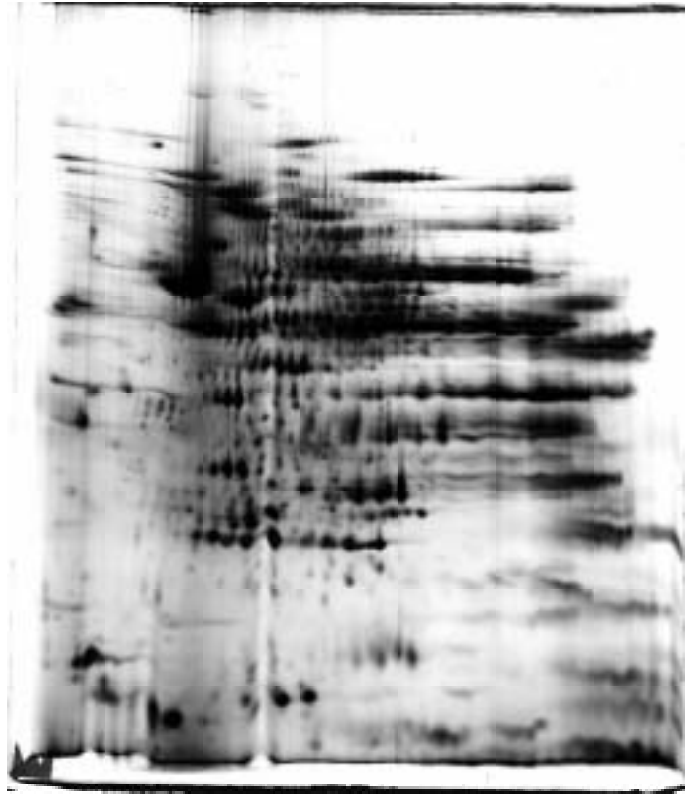

A  B
C  D

Figure 1

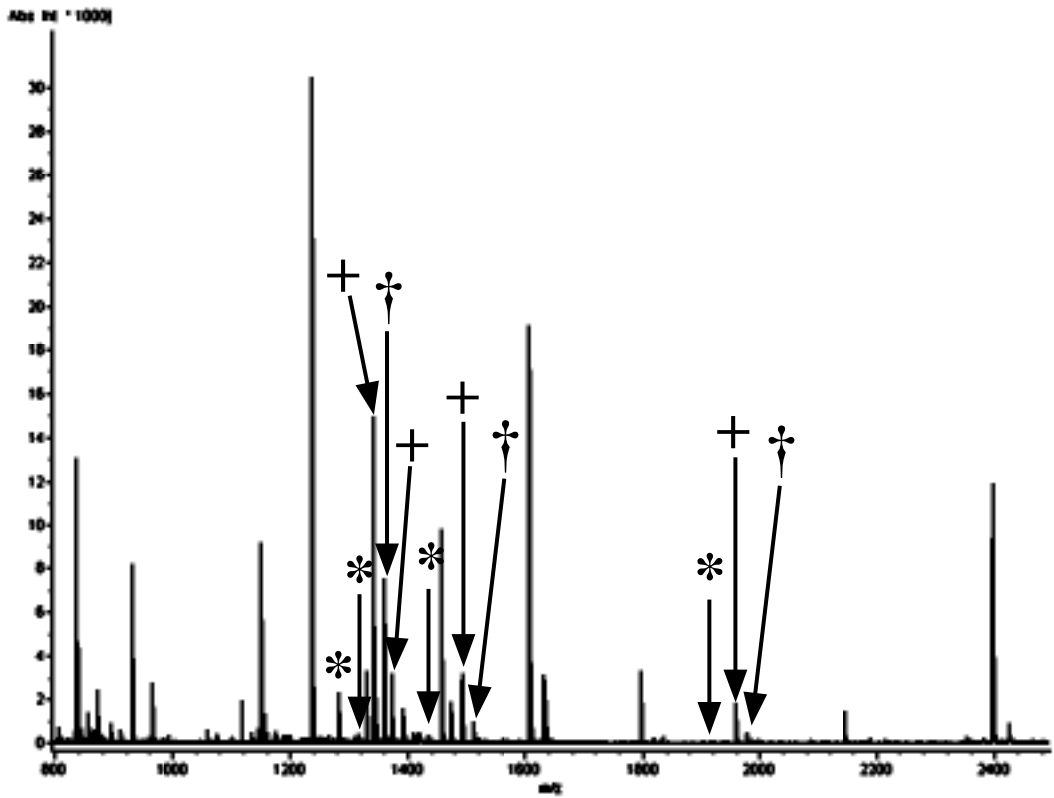

A

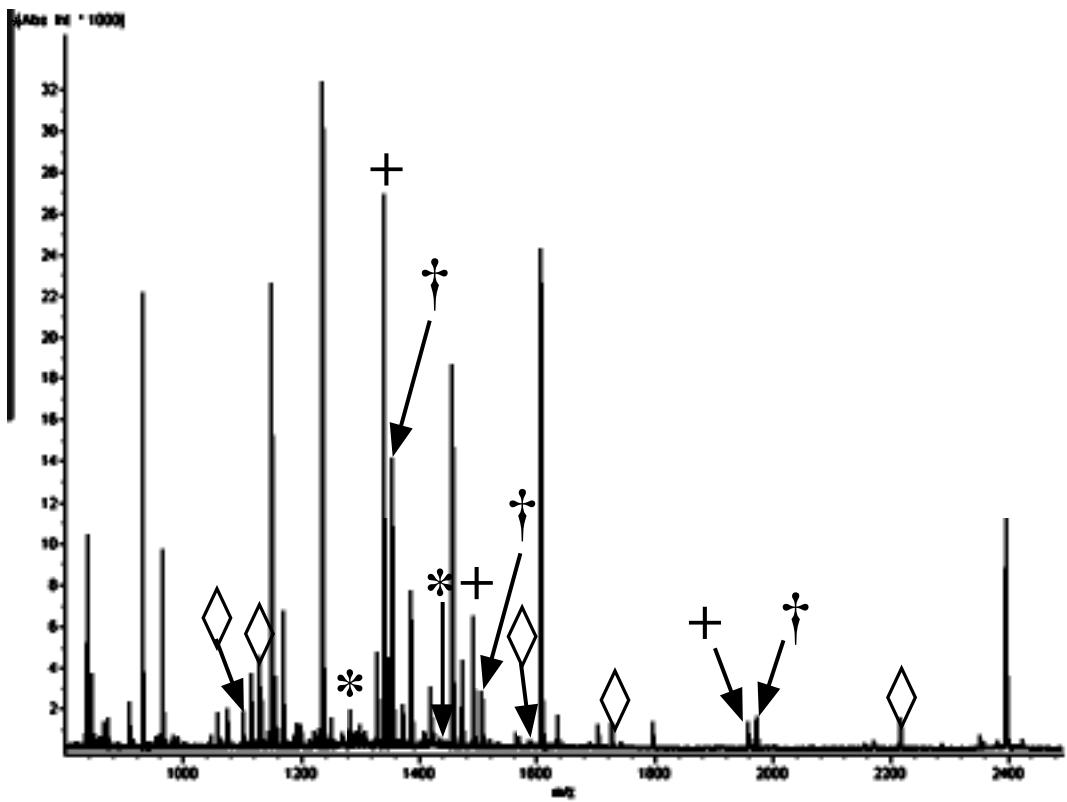

B

Figure 2F

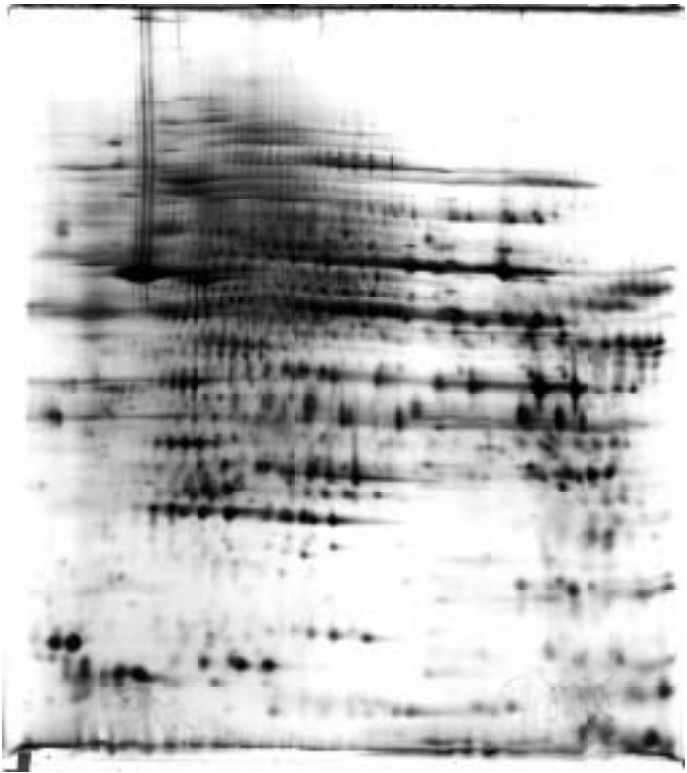

A

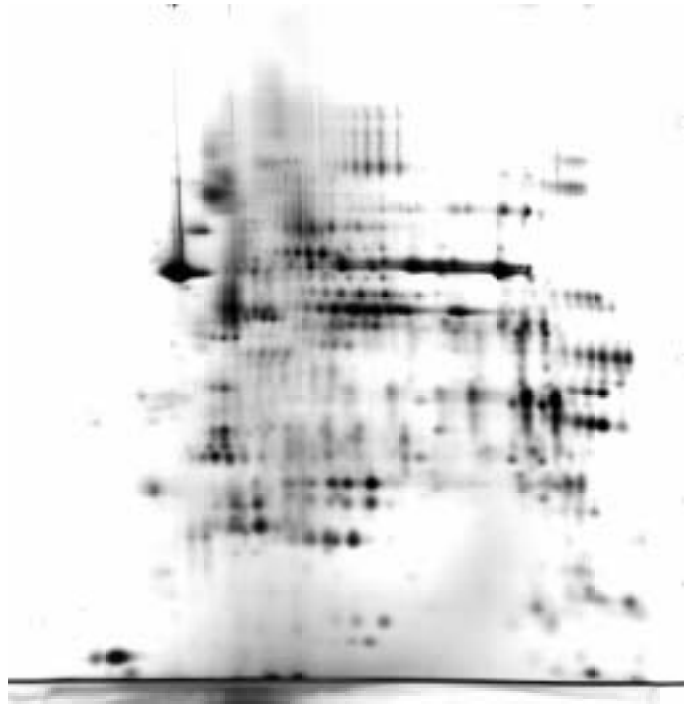

B

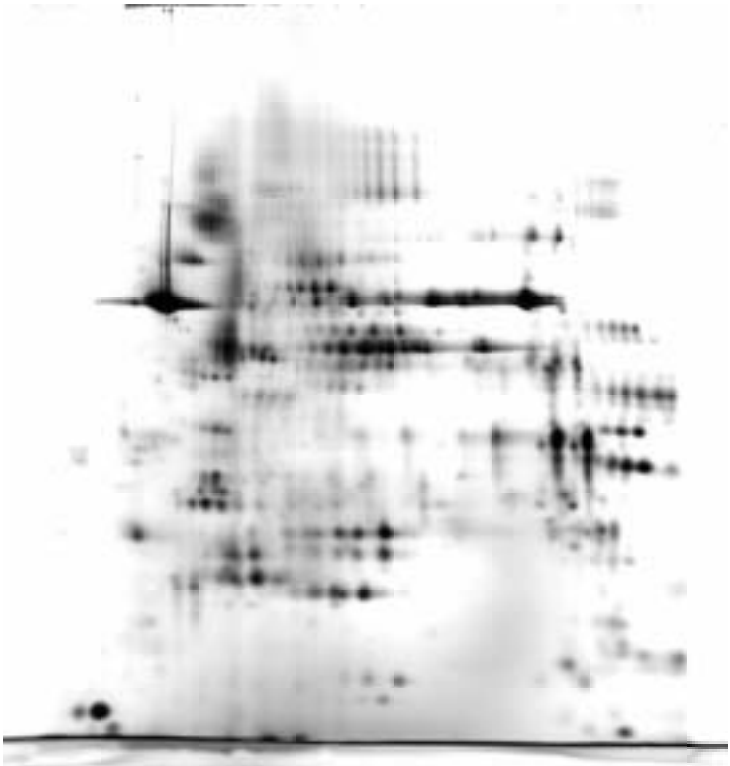

C

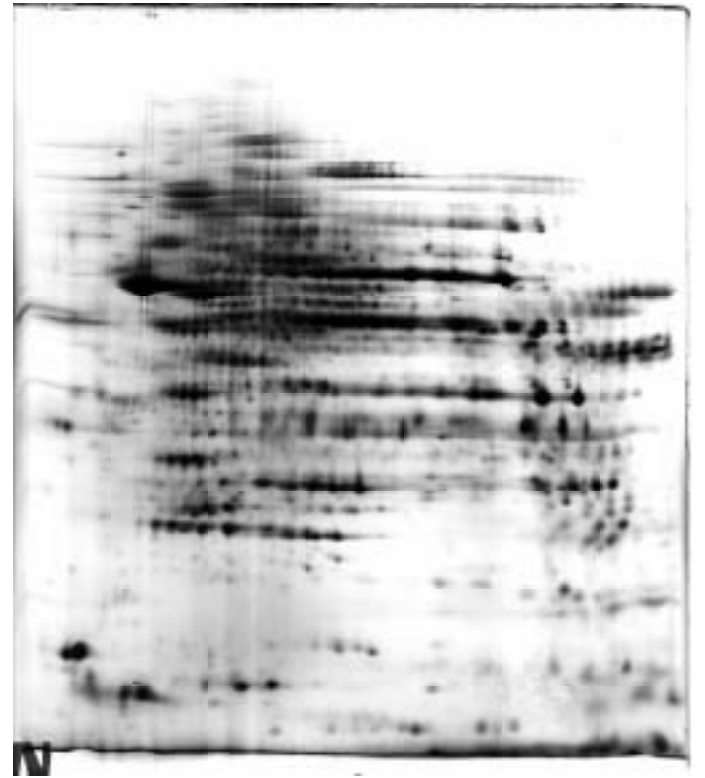

D

Figure 3

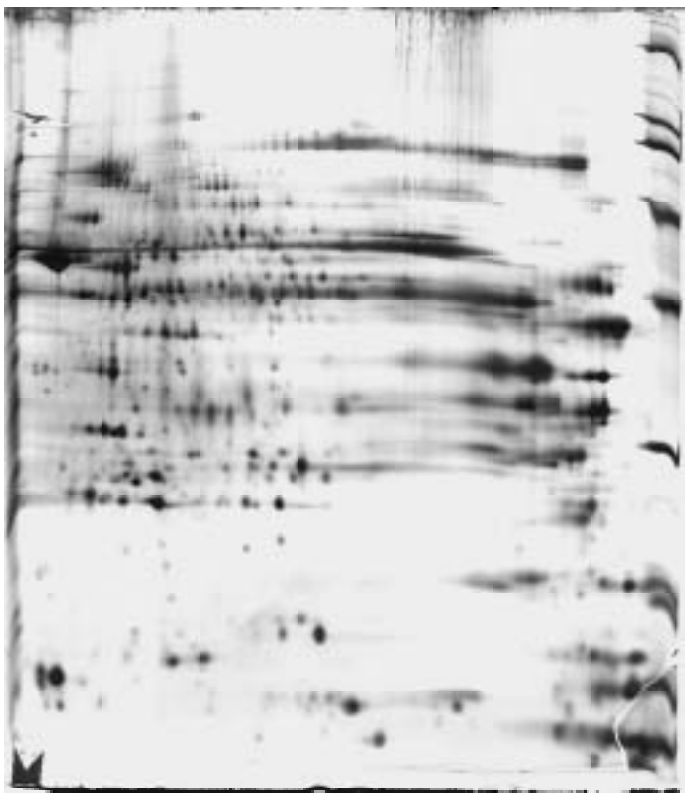 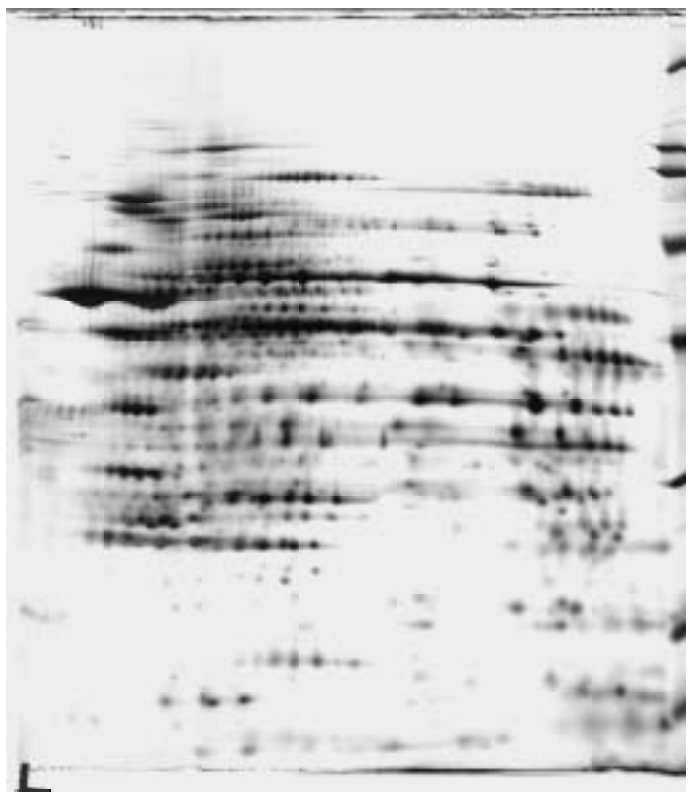

A                                B

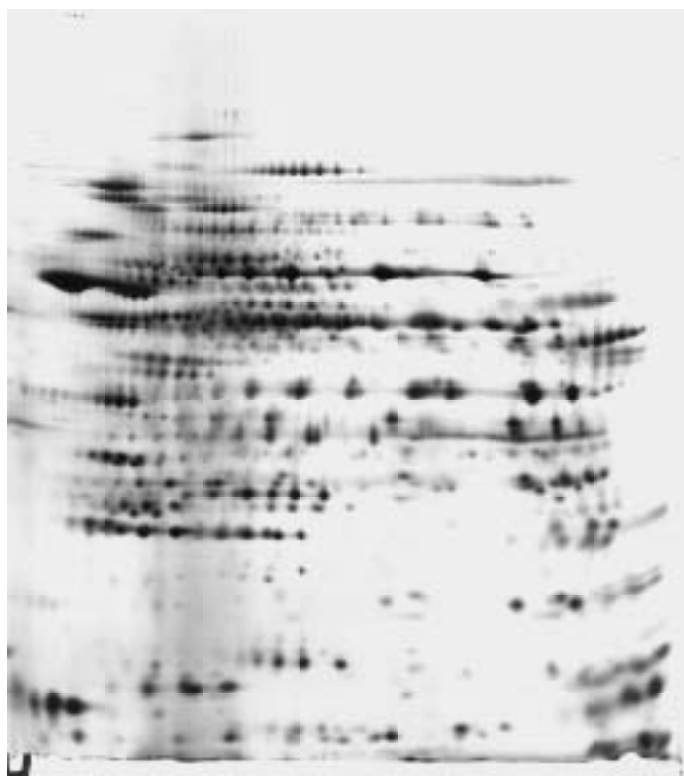 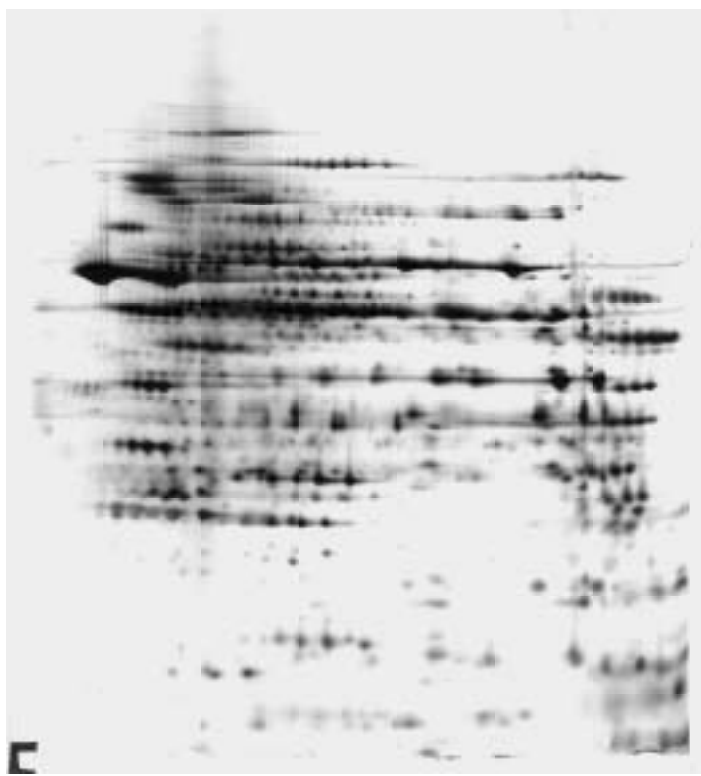

C                                D

Figure 4

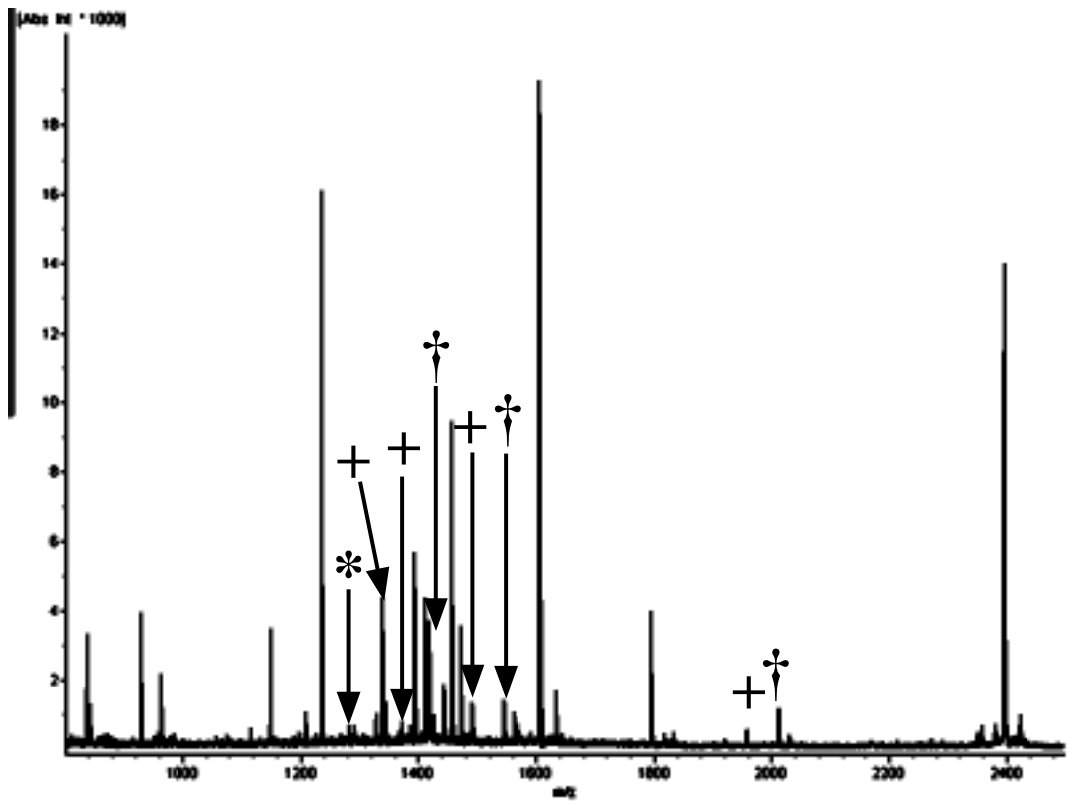
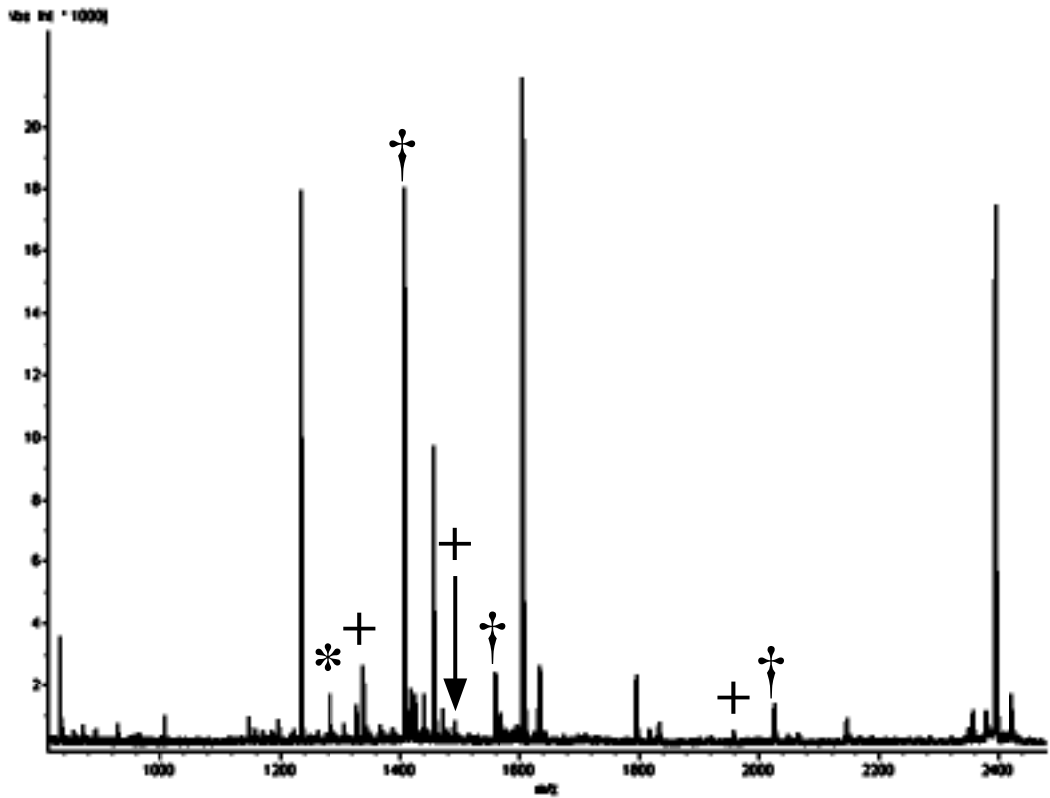

Figure 5

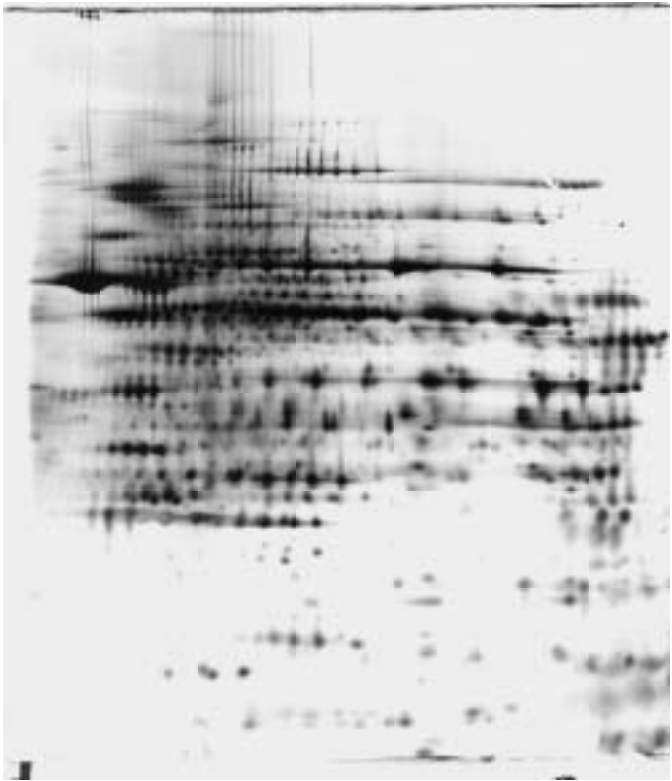
A

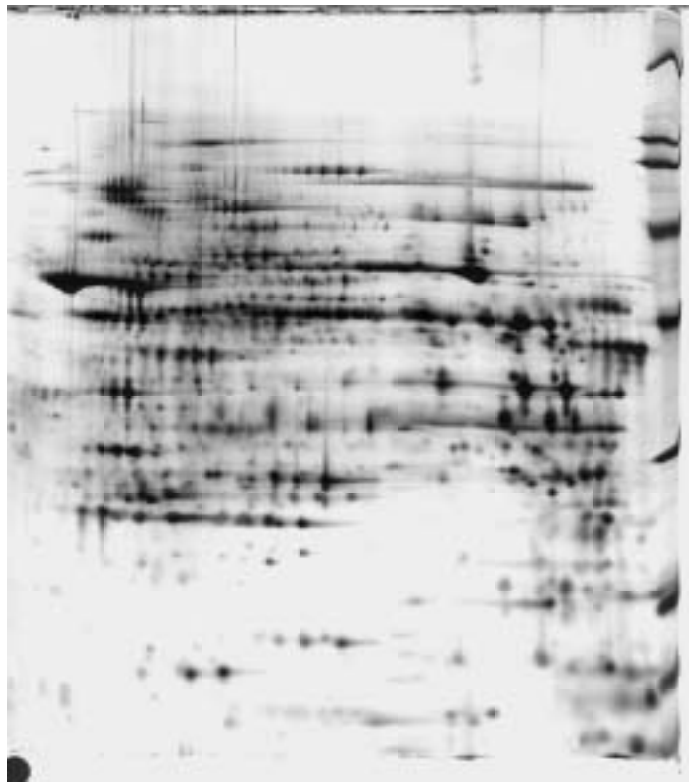
B

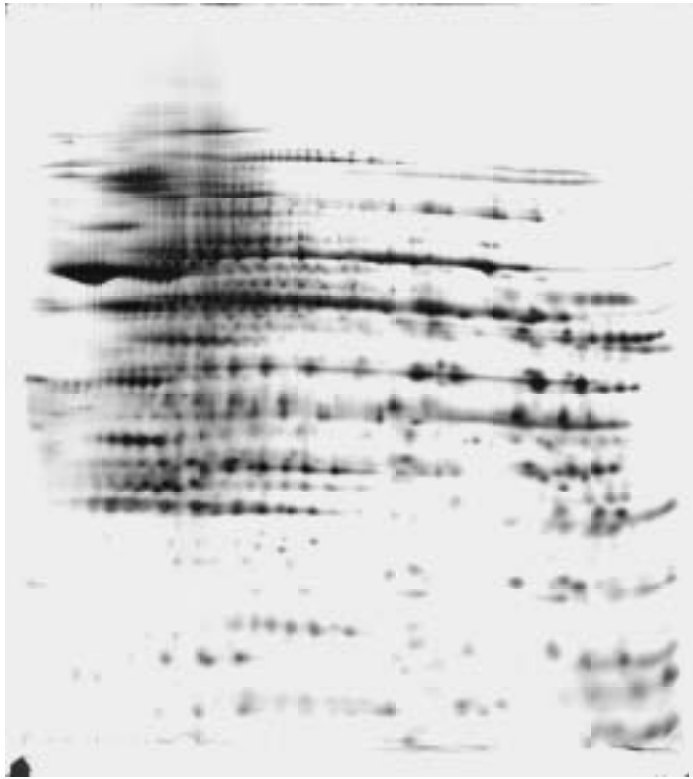
C

Figure 6

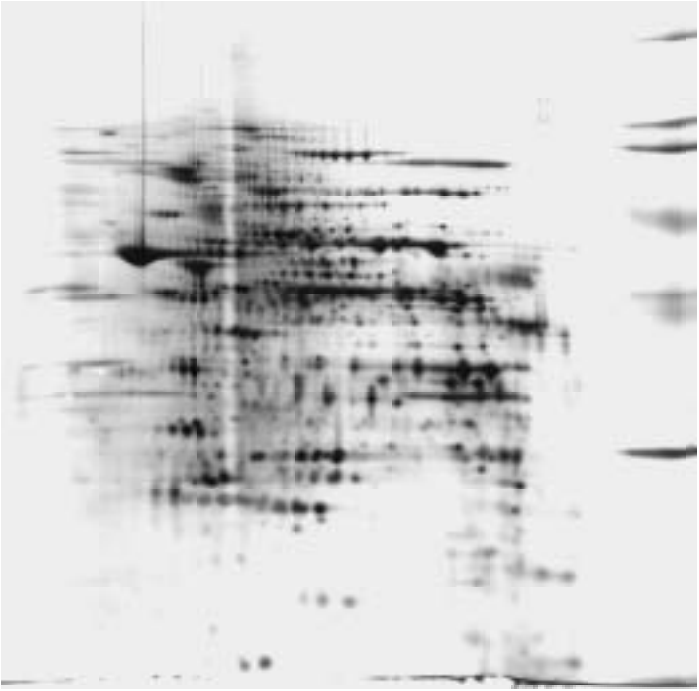 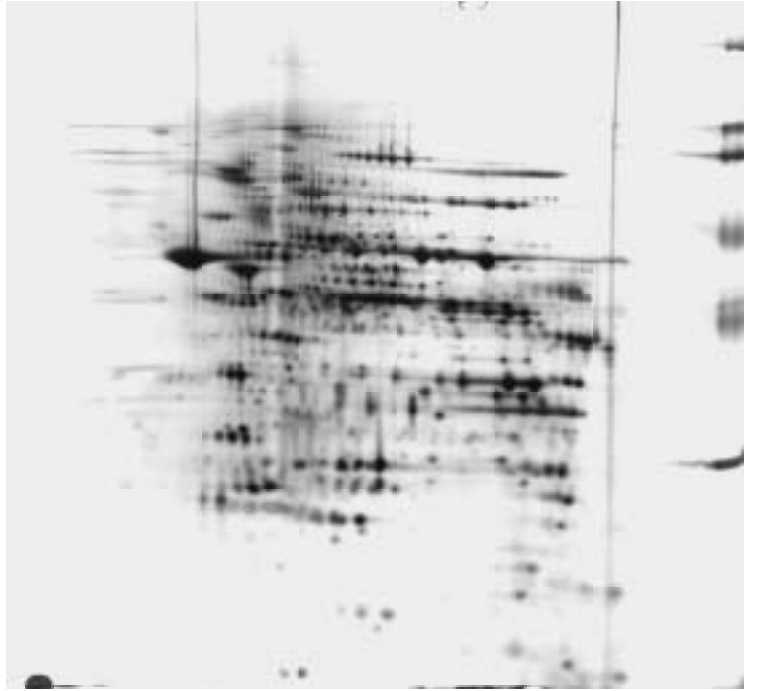

**A**  **B**

Figure 7

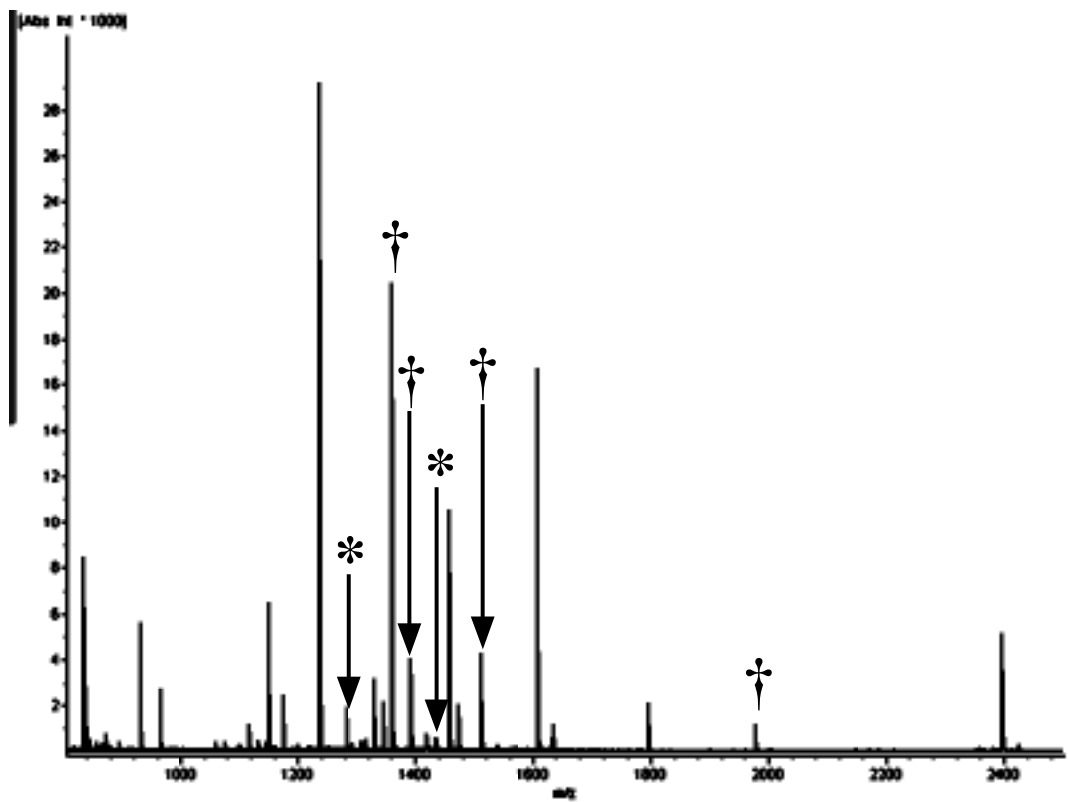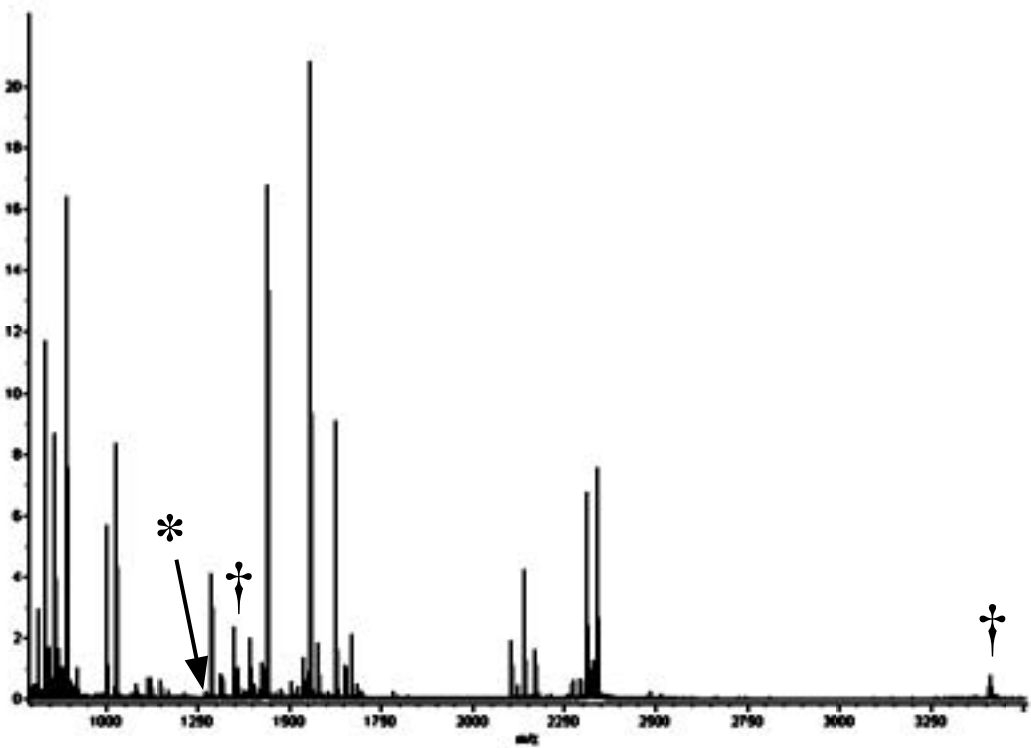

Figure 8

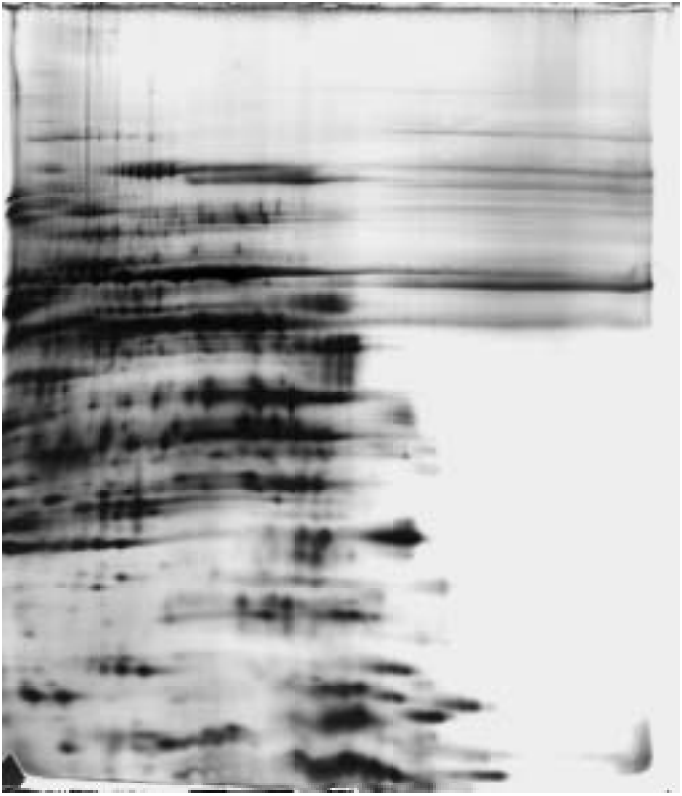 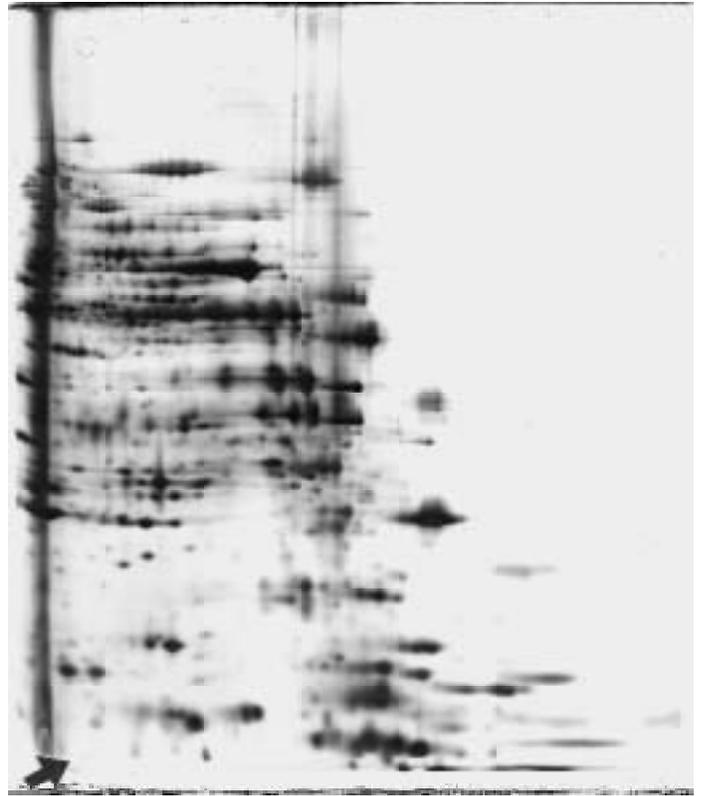

**A**  **B**

Figure 9